\def\1{\'{\i}}
\def\2{\~n}

\def\ep{\epsilon}

\def\ba{\begin{eqnarray}}
\def\ea{\end{eqnarray}}
\def\be{\begin{equation}}
\def\ee{\end{equation}}
\def\beq{\begin{equation}}
\def\eeq{\end{equation}}

\documentclass[10pt]{iopart}
\usepackage{iopams}
\usepackage{times}
\usepackage{cite}
\begin{document}

\title[]{On the new translational shape invariant potentials}
\author{Arturo Ramos
}

\address{
Departamento de An\'alisis Econ\'omico,
Universidad de Zaragoza, \\ Gran V\'{\i}a 2, E-50005 Zaragoza, Spain}

\address{\ E-mail: aramos@unizar.es}

\begin{abstract}
Recently, several authors have found new
translational shape invariant potentials not present
in classic classifications like that of Infeld and Hull.
For example, Quesne on the one hand and Bougie, Gangopadhyaya
and Mallow on the other have provided examples of them,
consisting on deformations of the classical ones.
We analyze the basic properties of the new examples
and observe a
compatibility equation which has to be satisfied by them.
We study particular cases of such equation and
give more examples of new translational shape invariant
potentials.
\end{abstract}

\pacs{03.65.-w,03.65.Fd}

\ams{81Q05,81Q60}




\section{Introduction}

In the late fourties and early fifties of last century,
Infeld and Hull \cite{Inf41,HulInf48,InfHul51}
introduced the so called \emph{factorization method},
for defining and solving problems in unidimensional
Quantum Mechanics. In the mid-eighties of last century
emerged the notion of \emph{shape invariance}
in Quantum Mechanics \cite{Gen83,GenKriv85}
which has a close resemblance to that of the factorization method.
In fact, it has been proved in a review article \cite{CarRam00}
the complete equivalence of both approaches.
As a result, a list of shape invariant potentials
have been produced, see, e.g., \cite{CarRam00,CooKhaSuk01}.
This list has been generalized to shape invariant potentials
which depend on $n>1$ parameters transformed by
translation \cite{CarRam00b}.

However, very recently, new shape invariant potentials
have been discovered whose parameter transforms by translation
and are not present in the classifications mentioned above.
The first of such examples is given by Quesne \cite{Que08},
work which inspired the research of Bougie, Gangopadhyaya
and Mallow \cite{BouGanMal10,BouGanMal11}.
More examples are found
in the work by Odake and Sasaki \cite{OdaSas09,OdaSas10}
although we will not consider
them in this article.
Quesne asks herself in \cite{Que08} about the reason why her
example is isospectral to a shape invariant potential
of the ordinary type (ordinary in the sense that it
belongs to the classical classifications already mentioned).

This paper studies the basic properties of the found examples
and apply the results in the search of new translational shape
invariant potentials using as data the classical shape invariant potentials
as classified in \cite{CarRam00}, which are essentially the same potentials
as in the classical Infeld and Hull \cite{InfHul51} classification.
The article is organized as follows. In the first section we recall
the classic framework of shape invariance. In the second
we describe the equations which satisfy the new examples of \cite{Que08,BouGanMal11} in order to induce a common framework
for these cases. We also particularize to the specific forms of
superpotentials given in \cite{CarRam00}. In the third section we
describe all the examples we have found with this technique.
In the fourth section we offer some conclusions and an
outlook for future research.

\section{Intertwinned Hamiltonians and Shape Invariance\label{ihsi}}

The simplest way of generating an exactly solvable
Hamiltonian $\widetilde H$ from a known one $H$ is just to
consider an invertible bounded operator $B$, with bounded inverse,
and defining $\widetilde H=BHB^{-1}$. This
transformed Hamiltonian $\widetilde H$ has the same spectrum as the starting one
$H$. As a generalization (see, e.g., \cite{CarMarPerRan98}), we will say
that two Hamiltonians $H$ and $\widetilde H$ are intertwined or
$A$-related when $AH=\widetilde HA$, where $A$ may have no inverse.
In this case, if $\psi $ is an eigenvector of
$H$ corresponding to the eigenvalue $\ep$ and $A\psi\neq 0$, at least
formally $A\psi$ is an eigenvector of $\widetilde H$ corresponding to the
same eigenvalue $\ep$.

If $A$ is a first order differential operator,
\ba
A=\frac {d}{dx}+W(x)\ ,\quad\quad\mbox{and}\quad\quad
A^{\dagger}=-\frac{d}{dx}+W(x)\ ,
\label{defAAdag}
\ea
then the relation $AH=\widetilde HA$, with
\begin{equation}
H=-\frac{d^2}{dx^2}+V(x)\,,\qquad \widetilde H=-\frac{d^2}{dx^2}+\widetilde V(x)\ ,
\label{defHHtil}
\end{equation}
leads to
$$
V=-2W'+\widetilde V, \qquad W(V-\widetilde V)=-W''-V'\ .
$$
Taking into account the first equation, the second becomes $2WW'=W''+V'$,
which can easily be integrated giving
\begin{equation}
V=W^2-W' + \ep\,,         \label{ricV}
\end{equation}
and then,
\begin{equation}
\widetilde V=W^2+W' + \ep\,,     \label{ricVtil}
\end{equation}
where $\ep$ is an integration constant. The important point here is that
$H$ and $\widetilde H$, given by (\ref{defHHtil}), are related by a first order
differential operator $A$, given by (\ref{defAAdag}), if and only if there
exist a constant $\ep$ and a function $W$ such that the pair of Riccati
equations (\ref{ricV}) and (\ref{ricVtil}) are satisfied
\emph{simultaneously}.  Moreover, this means that
both Hamiltonians can be factorized as
\begin{equation}
H=A^{\dag}A+\ep\,,\qquad \widetilde H=AA^{\dag}+\ep\ .\label{factorHHtil}
\end{equation}

Adding and subtracting equations (\ref{ricV}) and (\ref{ricVtil})
we obtain the equivalent pair which relates $V$ and $\widetilde V$
\begin{eqnarray}
\widetilde V-\ep&=&-(V-\ep)+2W^2\,,       \label{relVVtilcuad} \\
\widetilde V&=&V+2 W'\,.              \label{relVVtilder}
\end{eqnarray}
The function $W$ satisfying these equations is usually called the
\emph{superpotential}, the constant $\ep$ is the \emph{factorization energy}
or \emph{factorization constant} and $\widetilde V$ and $V$ (resp.
$\widetilde H$ and $H$) are said to be \emph{partner} potentials (resp. Hamiltonians).

Notice that the initial solvable Hamiltonian can arbitrarily be chosen as
$H$ or $\widetilde H$. In both cases the point will be to find a solution $W$ of
the corresponding Riccati equation (\ref{ricV}) or (\ref{ricVtil}) for a
specific factorization energy $\ep$. {}From this solution the expression for
the (possibly) new potential follows immediately from (\ref{relVVtilder}).

Note that these equations have an intimate relation with
what it is currently known as \emph{Darboux transformations} of
linear second-order differential equations \cite{Cru55,Inc56}, or in the context
of one-dimensional (or supersymmetric)
quantum mechanics \cite{MatSal91}.
In fact, it is easy to prove that the
equation (\ref{ricV}) can be transformed into
a Schr\"odinger equation $-\phi^{\prime\prime}+(V(x)-\ep)\phi=0$
by means of the change $-\phi^{\prime}/\phi=W$.
Likewise, by means of $\tilde\phi^{\prime}/\tilde\phi=W$
(\ref{ricVtil}) transforms into
$-\tilde\phi^{\prime\prime}+(\tilde V(x)-\ep)\tilde\phi=0$.
The relation between $V$ and $\tilde V$ is given by (\ref{relVVtilder}).
Obviously, $\phi \tilde\phi=1$, up to a non-vanishing constant factor.
It is also worth noting that these Schr\"odinger equations
express that $\phi$ and $\tilde\phi$ are respective eigenfunctions
of the Hamiltonians (\ref{defHHtil}) for the eigenvalue $\ep$.

The factorization method has been introduced by Infeld and Hull \cite{Inf41,HulInf48,InfHul51} providing the tools for
solving algebraically a class of unidimensional potentials
which showed a translational symmetry in one parameter.
In the mid-eighties of last century emerged the notion
of \emph{shape invariance} in Quantum Mechanics \cite{Gen83,GenKriv85}
which generalized the parameter space which is subject
to a invertible transformation.
Later, taken into account this generalization it has been proved that
both approaches coincide, see, e.g., \cite{CarRam00}.

In essence, it is taken equations (\ref{ricV}) and (\ref{ricVtil})
as a definition of the functions $V$, $\widetilde V$ in terms of
the function $W$ and some constant $\ep$.
After, one can assume that $W$ did depend on certain
set of parameters $a$, i.e., $W=W(x,a)$, and as a
consequence $V=V(x,a)$ and $\widetilde V=\widetilde V(x,a)$ as well.
Then, the necessary condition for $\widetilde V(x,a)$ to be
essentially of the same form as $V(x,a)$, maybe for a different choice
of the values of the parameters involved in $V$,
is known as shape invariance.
It amounts to assume the further relation
between $V(x,a)$ and $\widetilde V(x,a)$
\begin{equation}
\widetilde V(x,a)=V(x,f(a))+R(f(a))\,,\label{SIcond}
\end{equation}
where $f$ is an (invertible) transformation on the
parameter space\ $a$ and $R$ is some function of the parameters only.

Let us remark that it is the choice of the parameter space $a$
and of the (invertible) transformations $f(a)$ what define
the different types of shape invariant potentials.
Note that in principle, different types of shape invariant
potentials may have members in common.
Note as well that the function $f$ may be even
the identity, i.e., $f(a)=a$ for all $a$ \cite{AndCanIofNis00}.

Just writing the $a$-dependence
the equations (\ref{ricV}), (\ref{ricVtil}) become
\ba
V(x,a)-\ep&=&W^2-W'\,,            \label{ricVSI}  \\
\tilde V(x,a)-\ep&=&W^2+W'\,.       \label{ricVtilSI}
\ea
The simplest way of satisfying these equations
is assuming that $V(x,a)$ and $\tilde V(x,a)$ are
obtained from a superpotential function $W(x,a)$ by means of
\ba
V(x,a)-\ep&=&W^2(x,a)-W'(x,a)\,,                  \label{ricVSIsp}  \\
\tilde V(x,a)-\ep&=&W^2(x,a)+W'(x,a)\,.             \label{ricVtilSIsp}
\ea
The shape invariance property
requires the further condition (\ref{SIcond}) to be
satisfied, which in these terms reads
\begin{equation}
W^2(x,a)-W^2(x,f(a))+W'(x,f(a))+W'(x,a)=R(f(a))\ .
\label{SIsp}
\end{equation}
In practice, when searching shape invariant
potentials with a given parameter space $a$ and the
transformation function $f$, what it is done is to (try to) find
solutions for $W(x,a)$ and $R(a)$ of (\ref{SIsp}), instead of
solving the pair (\ref{ricVSIsp}), (\ref{ricVtilSIsp}) and then
imposing (\ref{SIcond}).

Now, we will consider the simplest but
particularly important case of shape invariant potentials
having only one parameter whose transformation law is a translation.
In other words, this case corresponds to the whole family of
factorizable problems treated in \cite{InfHul51}.
Thus, we will consider problems where the parameter
space is unidimensional, and the transformation law is
\begin{equation}
f(a)=a-\delta\,,\quad\quad\mbox{or}\quad\quad f(a)=a+\delta\,,
\label{a_tras}
\end{equation}
where $\delta\neq 0$.
In both cases we can normalize the parameter in units of $\delta$,
introducing the new parameter
\begin{equation}
m=\frac{a}{\delta}\,,\quad\quad\mbox{or}\quad\quad m=-\frac{a}{\delta}\,,
\label{m_norm}
\end{equation}
respectively. In each of these two possibilities
the transformation law reads, with a slight abuse of the notation $f$,
\begin{equation}
f(m)=m-1\,.
\label{m_tras}
\end{equation}
and the equations which should be solved in order
to find potentials in this class are
\ba
V(x,m)-\ep&=&W^2(x,m)-W'(x,m)\,,                  \label{ricVSI1p}        \\
\tilde V(x,m)-\ep&=&W^2(x,m)+W'(x,m)\,,             \label{ricVtilSI1p}
\ea
or the equivalent equations
\ba
\tilde V(x,m)-\ep&=&-(V(x,m)-\ep)+2\,W^2(x,m)\,,      \label{relVVtilcuadSI1p} \\
\tilde V(x,m)&=&V(x,m)+2\,W'(x,m)\,,              \label{relVVtilderSI1p}
\ea
as well as the shape invariance condition
\begin{equation}
\tilde V(x,m)=V(x,m-1)+R(m-1)\,.
\label{SIGed1p}
\end{equation}

\section{Properties of the new translational shape invariant potentials\label{prontsi}}

In the examples of \cite{Que08} and \cite{BouGanMal11}
the superpotential function takes the form of
\be
W(x,m)=W_0(x,m)+W_{1+}(x,m)-W_{1-}(x,m)\,, \label{Wgen}
\ee
where $W_0(x,m)$ is the superpotential of a pair of
shape invariant partner potentials of the classical type,
and $W_{1+}(x,m)$, $W_{1-}(x,m)$ are functions
of a type described below. Substituting (\ref{Wgen}) into
(\ref{ricVSI1p}) and (\ref{ricVtilSI1p}) it is observed that
the final partner potentials have the form
(the constant $\ep$ is taken as zero)
\ba
V(x,m)&=&V_0(x,m)-2W_{1+}'(x,m)\,,                  \label{Vg}        \\
\widetilde V(x,m)&=&\widetilde V_0(x,m)-2W_{1-}'(x,m)\,,    \label{tilVg}
\ea
where $V_0(x,m)$, $\widetilde V_0(x,m)$ is the pair of shape invariant partner potentials associated to $W_0(x,m)$. To this end, it is necessary that the following \emph{compatibility condition} hold:
\begin{eqnarray}
& & W_{1+}^2+W_{1+}^{\prime}+W_{1-}^2+W_{1-}^{\prime} \nonumber\\
& &\quad-2W_0W_{1-}+2W_0W_{1+}-2W_{1-}W_{1+}=0 \label{cc1}
\end{eqnarray}
(the dependence on the arguments has been omitted for brevity and clearness).
The condition of shape invariance (\ref{SIGed1p}) reads in this case
\begin{eqnarray}
& &\widetilde V(x,m)-V(x,m-1)-R(m-1)\nonumber\\
& &\quad=\widetilde V_{0}(x,m)-V_{0}(x,m-1)-R(m-1)\nonumber\\
& &\quad\quad-2W_{1-}^\prime(x,m)+2W_{1+}^\prime(x,m-1)=0\nonumber
\end{eqnarray}
that is equal, using (\ref{SIGed1p}) for the partner potentials
$\widetilde V_{0}(x,m)$ and $V_{0}(x,m)$ , to
$$
-2W_{1-}^\prime(x,m)+2W_{1+}^\prime(x,m-1)=0\,,
$$
thus we obtain a
second \emph{shape invariance condition}:
\begin{equation}
W_{1-}^\prime(x,m)=W_{1+}^\prime(x,m-1) \label{sic2}
\end{equation}

\subsection{Differential equation satisfied by $W_{1+}(x,m)$ and $W_{1-}(x,m)$\label{difeqapam}}

We will consider in this subsection superpotentials $W_0(x,m)$
of the form $W_0(x,m)=k_0(x)+m k_1(x)$, where $k_0(x)$ and $k_1(x)$
are not functions of $m$ and has been of use in the classification of shape invariant potentials, see \cite{CarRam00,InfHul51}.

It is further observed that in the examples of
\cite{Que08,BouGanMal11} the functions
$W_{1+}(x,m)$ and $W_{1-}(x,m)$ (in the notation of (\ref{Wgen}))
satisfy the differential equation
\be
W^\prime+W^2-k_1(x)W=0\,, \label{eqBern}
\ee
which is a Bernoulli equation with $n=1$
(it can be regarded as a special type
of Riccati equation, see, e.g.,\cite{CarRam99}).
Equation (\ref{eqBern}) is explicitly solvable
by two quadratures (we make two integration constants explicit):
\be
W(x)=\frac{d}{dx}\log\left(c_2+\int^x \exp\left(\int^y k_1(z)\,dz+\log(c_1)\right)\,dy\right)\,.\label{sgBer}
\ee

Using that $W_{1+}(x,m)$ and $W_{1-}(x,m)$
satisfy (\ref{eqBern}) in this case,
the compatibility condition (\ref{cc1})
reduces to
\begin{eqnarray}
2(k_0+m k_1)(W_{1+}-W_{1-})+k_1(W_{1+}+W_{1-})-2W_{1-}W_{1+}=0\,, \label{cc2}
\end{eqnarray}
which is an algebraic equation to be satisfied.
The method for obtaining new translational
shape invariant potentials is now clear: to select solutions
of (\ref{eqBern}) which satisfy (\ref{cc2}) for each specific
case of superpotential $k_0(x)+m k_1(x)$. We do it in the next section.

\section{Examples\label{examples}}

We will analyze, following the previous procedure, all cases of
superpotential\  $k_0(x)+m k_1(x)$ present in \cite{CarRam00,InfHul51}.
We will obtain therefore, the potentials
of \cite{Que08,BouGanMal11} (slightly generalized) and other new potentials.

\subsection{Case of $k_0(x)+m k_1(x)
=\frac b c\tanh(c x)+\frac{d}{\cosh(c x)}+mc\tanh(c x)$}

For this case $x\in(-\infty,\infty)$ and
$k_1(x)=c \tanh(c x)$, where $c>0$ is a constant.
Then, the solutions $W_{1+}(x,m)$ and $W_{1-}(x,m)$ to
(\ref{eqBern}) can be written
\ba
W_{1+}(x)&=&\frac{c c_1 \cosh(c x)}{c c_2+c_1 \sinh(c x)} \nonumber\\
W_{1-}(x)&=&\frac{c c_3 \cosh(c x)}{c c_4+c_3 \sinh(c x)} \nonumber
\ea
where $c_1,c_2,c_3,c_4$ are constants to be determined.
Inserting these expressions into (\ref{cc2}) lead to the following
relations between the previous constants:
\ba
c_2&=&\frac{c_1 \left(2 b+2c^2 m+c^2\right)}{2 c^2d}\nonumber\\
c_4&=&\frac{c_3 \left(2 b+2c^2 m-c^2\right)}{2 c^2d}\nonumber
\ea
such that $W_{1+}(x,m)$ and $W_{1-}(x,m)$ become
\ba
W_{1+}(x,m)&=&\frac{2 c^2 d \cosh(c x)}{2b+c^2(2m+1)+2 c d\sinh(c x)} \nonumber\\
W_{1-}(x,m)&=&\frac{2 c^2 d \cosh(c x)}{2b+c^2(2m-1)+2 c d\sinh(c x)} \nonumber
\ea
This case generalize slightly that
found in \cite{BouGanMal11}, with a different notation.
However, it has a substantial defect, which is
that both of $W_{1+}(x,m)$ and $W_{1-}(x,m)$ present
singularities at some point of the domain $(-\infty,\infty)$.
The reason is that when $d\neq 0$ the
function $2 c d\sinh(c x)$ has
the whole real line as image and is strictly monotone,
and the graph of vertical translations of it always
crosses the horizontal axis once.

\subsection{Case of $k_0(x)+m k_1(x)
=\frac b c\coth(c x)+\frac{d}{\sinh(c x)}+m c\coth(c x)$}

For this case $x\in(0,\infty)$ and  $k_1(x)=c \coth(c x)$,
where $c>0$ is a constant.
Following the procedure analogous to the previous subsection
we find that $W_{1+}(x,m)$ and $W_{1-}(x,m)$ become
\ba
W_{1+}(x,m)&=&\frac{2 c^2 d \sinh(c x)}{2 c d\cosh(c x)-2b-c^2(2m+1)} \nonumber\\
W_{1-}(x,m)&=&\frac{2 c^2 d \sinh(c x)}{2 c d\cosh(c x)-2b+c^2(1-2m)} \nonumber
\ea
These functions are free of singularities, when $d>0$ if $m<\frac{2cd-2b-c^2}{2c^2}$, and when $d<0$ if $m>\frac{2cd-2b+c^2}{2c^2}$.
This leads to a new case in the literature.

\subsection{Case of $k_0(x)+m k_1(x)=\pm\frac b c+d \exp(\mp c x)\pm m c $}

For this case $x\in(-\infty,\infty)$ and $k_1(x)=\pm c$,
where $c>0$ is a constant.
Following the procedure analogous to the previous subsections
we find that $W_{1+}(x,m)$ and $W_{1-}(x,m)$ become
\ba
W_{1+}(x,m)&=&\pm c\nonumber\\
W_{1-}(x,m)&=&\pm c\nonumber
\ea
This leads to a new, although rather trivial, case in the literature.

\subsection{Case of $k_0(x)+m k_1(x)=\frac b 2 x+\frac d x+\frac m x$}

For this case $x\in(0,\infty)$ and $k_1(x)=\frac 1 x$.
Following the procedure analogous to the previous subsections
we find that $W_{1+}(x,m)$ and $W_{1-}(x,m)$ become
\ba
W_{1+}(x,m)&=&\frac{2 b x}{b x^2-1-2 d-2 m}   \nonumber\\
W_{1-}(x,m)&=&\frac{2 b x}{b x^2+1-2 d-2 m}    \nonumber
\ea
These functions are free of singularities when $m<-\frac 1 2(1+2d)$.
This is the case of \cite{Que08} although slightly generalized.

\subsection{Case of $k_0(x)+m k_1(x)=b x+d$}

For this case $x\in(-\infty,\infty)$ and $k_1(x)=0$.
Following the procedure analogous to the previous subsections
we find no nontrivial solutions for  $W_{1+}(x,m)$ and $W_{1-}(x,m)$.

\subsection{Case of $k_0(x)+m k_1(x)
=\frac b c\tan(c x)+\frac{d}{\cos(c x)}-m c\tan(c x)$}

For this case $x\in\left(-\frac{\pi}{2c},\frac{\pi}{2c}\right)$ and
$k_1(x)=-c \tan(c x)$, where $c>0$ is a constant.
Following the procedure analogous to the previous subsections
we find that $W_{1+}(x,m)$ and $W_{1-}(x,m)$ become
\ba
W_{1+}(x,m)&=&\frac{2 c^2 d \cos(c x)}{2 c d\sin(c x)+2b-c^2(2m+1)} \nonumber\\
W_{1-}(x,m)&=&\frac{2 c^2 d \cos(c x)}{2 c d\sin(c x)+2b+c^2(1-2m)} \nonumber
\ea
These functions are free of singularities, when $d>0$ if $m>\frac{2b+c^2+2cd}{2c^2}$ or $m<\frac{2b-c^2-2cd}{2c^2}$,
and when $d<0$ if $m>\frac{2b+c^2-2cd}{2c^2}$
or $m<\frac{2b-c^2+2cd}{2c^2}$.
This leads to a new case in the literature.

\subsection{Case of $k_0(x)+m k_1(x)
=-\frac b c\cot(c x)+\frac{d}{\sin(c x)}-m c\cot(c x)$}

For this case $x\in\left(0,\frac{\pi}{c}\right)$
and $k_1(x)=c \cot(c x)$, where $c>0$ is a constant.
Following the procedure analogous to the previous subsections
we find that $W_{1+}(x,m)$ and $W_{1-}(x,m)$ become
\ba
W_{1+}(x,m)&=&\frac{2 c^2 d \sin(c x)}{2b-c^2(2m+1)-2 c d\cos(c x)} \nonumber\\
W_{1-}(x,m)&=&\frac{2 c^2 d \sin(c x)}{2b+c^2(1-2m)-2 c d\cos(c x)} \nonumber
\ea
These functions are free of singularities, when $d>0$ if $m>\frac{2b+c^2+2cd}{2c^2}$ or $m<\frac{2b-c^2-2cd}{2c^2}$,
and when $d<0$ if $m>\frac{2b+c^2-2cd}{2c^2}$
or $m<\frac{2b-c^2+2cd}{2c^2}$.
This leads to a new case in the literature,
related to that of the previous case by a shift
in the variable\quad $x$ in $\frac{\pi}{2c}$.

\subsection{Case of $k_0(x)+m k_1(x)=\mp i\frac b c+d \exp(\mp i c x)\pm m i c $}

For this case $x\in(-\infty,\infty)$
and $k_1(x)=\pm i c$, where $c>0$ is a constant.
Following the procedure analogous to the previous subsections
we find that $W_{1+}(x,m)$ and $W_{1-}(x,m)$ become
\ba
W_{1+}(x,m)&=&\pm i c\nonumber\\
W_{1-}(x,m)&=&\pm i c\nonumber
\ea
This leads to a new, although also trivial, case in the literature.

This is to be remarked that in all these
cases $W_{1+}(x,m)$ and $W_{1-}(x,m)$ satisfy a condition slightly
stronger than (\ref{sic2}), namely
\begin{equation}
W_{1-}(x,m)=W_{1+}(x,m-1) \label{sic3}
\end{equation}
which obviously implies (\ref{sic2}). This means
that imposing (\ref{cc2}) is a stronger condition
than (\ref{sic2}) for the cases studied.

We also tried this procedure for the case of superpotential
$W_0(x,m)=q/m+m k_1(x)$, where $q$ is a real constant.
But we have obtained no nontrivial cases.

\section{Conclusions and outlook}

We have studied the properties of newly discovered
translational shape invariant potentials in the literature,
obtained by different means:
studying exceptional orthogonal polynomials in \cite{Que08},
and expansions in $\hbar$ in \cite{BouGanMal10,BouGanMal11}.
We have observed that they satisfy several equations and
set up a new approach based upon them. In fact, the form
of the final partner potentials obtained leads to the
fulfillment of a compatibility condition.

For the special case of $W_0(x,m)$ affine in $m$ it moreover
holds that the extra terms added to $W_0(x,m)$ satisfy each the same
Bernoulli equation, which is explicitly solvable. The compatibility
condition becomes no longer differential but an algebraic condition.
The constants of the solutions of the Bernoulli equation give enough
freedom in order to satisfy the (algebraic) compatibility condition.
We have found in this way the cases
of \cite{Que08,BouGanMal11}
and new ones.

For the special case of $W_0(x,m)=q/m+m k_1(x)$ no nontrivial
solutions can be found (with the assumptions of Section 4).
This is similar to what happened
to the extension to $n>1$ parameters transformed by translation
to this form of superpotential: no new nontrivial solutions
were found, see \cite{CarRam00b}.

Also, it would be interesting to see whether the present approach
applies to the translational shape invariant potentials found in
\cite{OdaSas09,OdaSas10}.

The form of the final potentials (\ref{Vg}) and (\ref{tilVg})
recalls the ordinary B\"acklund--Darboux transformations in
one-dimensional quantum mechanics, see \cite{Cru55,Inc56} for
a classic treatment and \cite{CarFerRam01,CarRam08} for
a geometric approach. However, the \emph{transforming function}
$W_{1-}(x,m)$ should satisfy a Riccati
equation of the type $W^\prime+W^2=\widetilde V_0$. Instead, it
satisfies a Bernoulli equation. Maybe the work
in \cite{CarRam00c} could help in understanding the problem.
This implies that the description
of the situation studied in this paper by means of Darboux
transformations is an open question, as it is the justification of the
isospectrality of the final potentials obtained with the initial
pair of partner potentials.

Finally, it would be of importance to determine whether the
compatibility condition (\ref{cc1}) admits more solutions, even
starting from potentials which do not conform a pair of shape
invariant potentials. For that, it might be useful to consider
cases where the extra terms in the superpotential do not satisfy
the previous Bernoulli equation but maybe other relations. This is
another open problem.

All of this will lead to interesting new research work.

\section*{Acknowledgements}
We acknowledge important remarks made by the referee. This work is
supported by Spanish Ministry of Science and Technology,
project ECO2009-09332 and by Aragon Government,
ADETRE Consolidated Group.

\end{document}